# Self-Similarity and Scaling in Forest Communities

Filippo Simini[1], Tommaso Anfodillo[2], Marco Carrer[2], Jayanth R. Banavar[3]

& Amos Maritan[1]

[1] *Dipartimento di Fisica ``G. Galilei'', Università di Padova, CNISM and INFN, via Marzolo 8, 35131 Padova, Italy.*

[2] *Dipartimento Territorio e Sistemi Agro-Forestali, Università di Padova, AGRIPOLIS, viale dell'Università 16, 35020 Legnaro (PD), Italy.*

[3] *Department of Physics, The Pennsylvania State University, 104 Davey Laboratory, University Park, Pennsylvania 16802.*

## Abstract

**Ecological communities exhibit pervasive patterns and inter-relationships between size, abundance, and the availability of resources. We use scaling ideas to develop a unified, model-independent framework for understanding the distribution of tree sizes, their energy use and spatial distribution in tropical forests. We demonstrate that the scaling of the tree crown at the individual level drives the forest structure when resources are fully used. Our predictions match perfectly with the scaling behaviour of an exactly solvable self-similar model of a forest and are in good accord with empirical data. The range, over which pure power law behaviour is observed, depends on the available amount of resources. The scaling framework can be used for assessing the effects of natural and anthropogenic disturbances on ecosystem structure and functionality.**

Understanding the interrelationships between patterns of size, abundance and resource availability in tree-dominated communities has proved to be a daunting challenge (*1-24*). Power law patterns are ubiquitous in plant communities but they are only observed up to a characteristic scale. For example, tree-size distributions in tropical forests scale as a pure power law only when trees smaller than ~ 30 - 40 cm in diameter are considered and there are fewer larger trees than predicted by a simple power law relationship, (*2, 4, 6, 9, 11-16, 18-23,25*). Recent comparative studies (*20, 21*) of the predictions of metabolic ecology (*4, 6, 7, 9, 12, 14, 15, 18, 19, 22, 23*) and demographic equilibrium theory (*16, 17, 21*) suggest the absence of a unique scaling relationship between tree abundance and size in tropical forests across the world. We demonstrate that one can use scaling (*26-30*) to disentangle universal power law behavior from community dependent detail. We show that scaling yields a *model-independent* description of ecological communities subject to resource limitations. We apply this general framework to the tree community in the Barro Colorado Island (BCI) forest (*25*) and show how scaling predicts relationships between ecological quantities in good accord with empirical data. Our predictions match perfectly with the scaling behavior of an exactly solvable self-similar model of a forest. While some of our results are in agreement with earlier work of Brown, Enquist, and West (*4, 6, 7, 9, 12, 14, 15, 18, 19, 22, 23*), others differ from earlier expectations. Our scaling analysis allows one to analyze situations in which one does not have pure power law behavior, determine the range over which power law behavior is present, and estimate the exponents.

As in earlier work (see, e.g. Ref. 22 and 23 and references therein), we assume that the forest is in steady state and we do not consider the distinction between different tree species. We base our derivations on the following four hypotheses, of which the first two concern a single tree whereas the last two pertain to the whole forest:

1. Tree shape: For a tree of height *h*, the transverse extension or crown radius is *postulated* to scale as $r_{cro} \sim h^H$. Thus, the crown volume scales as $h^{1+2H}$. Quite generally, $H \leq 1$. $H=1$ would imply an isometric tree shape, whereas $H<1$ would result in a taller tree being more elongated than a smaller one. The scaling analysis predicts the dependencies of various exponents characterizing the *individual tree and the forest* on this shape exponent *H* and thus provides links between exponents.

2. Energy optimization of a tree: The metabolic rate – mass relationship is obtainable by *maximizing the metabolic rate, B, for a given tree mass, M*. In agreement with empirical data, the metabolic rate, *B*, of a tree is assumed to be proportional to the number of leaves or to the tree crown volume, $B \sim h^{1+2H}$. This optimization is performed under the hypothesis that the average volume flow rate to the leaves is mass independent* (*31*).

3. Energy optimization of the forest: In order *to maximize the energy utilized by the forest, the leaves must fill the volume of the forest* which is proportional to the product of the forest area, *A*, [or equivalently the total number of trees in the forest (*10*)] and the typical height of the tallest trees (we denote this by $h_c$). We show below that this allows one to deduce that the tree height probability distribution function (PDF), $p_h(h)$, of a forest is a power law, when h < $h_c$, characterized by an exponent that depends on *H*.

4. Scaling: We generalize the pure power law behavior that we will deduce from the previous hypotheses building on the finite size scaling approach (*26-30*). The power of the scaling framework is that it will allow us to carry out a collapse (see Supporting Information) to deduce the range of parameters over which pure power law behaviour holds and determine the exponents. We consider a scaling form for the fraction of trees with height between $h$ and $h + dh$, given the typical height of the tallest trees in the forest is $h_c$, $p_h(h \mid h_c)dh = h^{-\alpha} f_h(h/h_c)dh$ (see Supporting Information), where $\alpha$ is the familiar power law exponent. $h_c$, as shown below, is a measure of the average resource use per tree or equivalently per unit area. The scaling function $f_h(h/h_c)$ is postulated to tend to a constant value when $h \ll h_c$, thus leading to pure power behaviour, $p_h(h|h_c) \sim h^{-\alpha}$, and approaches zero when $h$ approaches $h_c$ from below or is greater than it.

The scaling theory, based on the last hypothesis, thus takes into account the resource limitation in an ecological community, which cuts off pure power law behavior. While all scaling relationships involve a power law portion, the presence of the scaling function and the inherent characteristic height arising from resource limitations typically result in the power law behaviour occurring over a limited range of scales.

**Results**

We now proceed to deduce power law exponents and to a verification of the validity of finite size scaling (*29, 30*) with the available data from the BCI forest (*25*).

There are at least two distinct masses that one may attribute to a tree – the first is the mass of the fluid contained within the transportation network (i.e. the sapwood) and the second is the total tree mass including also the heartwood (which provides structural stability). The former mass is, of course, contained in the latter. We will assume, for simplicity, that the two masses scales isometrically, i.e. they are proportional to each other. Based on a general theorem pertaining to transportation networks (*5*), the maximum metabolic rate for a tree of mass $M$ is given by $B \sim M/h$ yielding $M \sim h^{2+2H}$ on using hypothesis 2. This result follows from the observation that efficient directed transport along the tree ensures that the mean distance from the source to all the leaves scales as $h$. Thus the metabolic rate-mass relationship takes the form $B \sim M^{(1+2H)/(2+2H)}$. We use the same optimization principle of maximizing the metabolic rate for a given mass to deduce that the optimal tree shape is characterized by $H = 1$, yielding the maximum value of the metabolic rate-mass exponent $(1+2H)/(2+2H)$ of $3/4$. This optimal case yields the classic Kleiber law (*1, 2*), $B \sim M^{3/4}$. Thus in the optimal case, $B \sim h^3$ and $M \sim h^4$. The total tree mass scales isometrically with the mass of the stem (*14*) and therefore $M \sim r^2 h$, where $r$ is the stem diameter. Thus $r \sim h^{3/2}$, coinciding with the result obtained from considerations of buckling (*1*). This result justifies the simplifying assumption made above of the isometric scaling of the two distinct definitions of tree mass $M$. This relationship between tree diameter and height predicts the tapering of the tree trunk and leads to the pleasing result that the metabolic rate $B \sim r^2$ as empirically demonstrated (11).

From allometric theory (*1, 2*), the characteristic biological time scales as $M/B$. It is the length of time required for a non-feeding animal to exhaust its stored metabolic energy or for blood circulation to take place in an organism. Thus, the characteristic mortality rate is predicted to be proportional to $B/M$ and scales as $h^{-1} \sim M^{-1/4} \sim r^{-2/3}$, which is in accord with the empirical data presented by Enquist et al. (*22*) when $H=1$. A refinement of the previous argument (see Supporting Information) allows one to bridge metabolic ecology (*4, 6, 7, 9, 12, 14, 15, 18, 19, 22, 23*) with demographic equilibrium theory (*16, 17, 21*).

To summarize, we have used a single optimization principle of maximizing the metabolic rate for a given mass to derive the shape and energy intake of a tree. Now, we utilize the *same principle at the level of a forest*, i.e. hypothesis 3, to determine the forest structure.

Let us first assume, consistent with hypothesis 4, that the PDF of the tree heights, $p_h(h)$, is zero both below some recruitment height, $h_0$, (lower cut-off) and above the typical height of the tallest tree, $h_c$, (upper cut-off) with $h_c \gg h_0$. Using hypothesis 3, the total energy utilized by the whole forest is given by the alternative expressions in the two sides of the following equation

$$A h_c = A \int_{h_0}^{h_c} dh\, p_h(h)\, h^{1+2H} \tag{1}$$

$A h_c$ is the total volume at disposal of the forest whereas $A\, dh\, p_h(h)$ is the total number of trees with heights in the interval $(h, h+dh)$. Thus $A\, dh\, p_h(h)\, h^{1+2H}$ is the metabolic rate of (or volume occupied by) trees with their heights in that range. If $p_h(h)$ is a continuous function, the above equation readily implies that

$$p_h(h) \stackrel{H=1}{\propto} h^{-3}\, \Theta(1 - h/h_c), \quad h > h_0 \tag{2}$$

where the $\Theta$ function is $1$ if its argument is positive and zero otherwise. The distribution of stem radii follows from the relationship between $r$ and $h$ to be $p_r(r) \propto r^{-7/3}\, \Theta(1 - r/r_c)$ with the cut-off value $r_c \sim h_c^{3/2}$. The case for generic $H$ is reported in Table 1 as well as the exponent values for the distribution of the crown height, the metabolic rate, the plant mass and other attributes using the standard rule for the change of variables. The scaling form postulated in hypothesis 4 for the tree height PDF (and the derived ones for other related variables) follows on substituting $\Theta(1-h/h_c)$ with a more general function $f_h(h/h_c)$.

The energy equivalence principle states that when trees are binned in discrete size classes, the total energy consumed within each class is the same. We find that this does hold when the size classes are based on tree height (see Supporting Information) and not on tree radius as has been suggested previously (*9, 22, 23*). The characteristic height of trees, $h_c$, is a measure of the average energy use per unit area or equivalently per tree. To summarize, we have used the same optimization principle at the tree level and at the forest level of maximizing the metabolic efficiency to derive relationships between, and the values of, exponents characterizing individual tree shape and forest structure. For the forest, the power law behavior of the distribution of tree heights or diameters is derived and not assumed *a priori*. However the range over which pure power law behavior is observed is limited. We turn now to the issue of how, in practice, one might determine the range over which pure power law behavior holds and estimate the values of exponents.

We begin by introducing a new variable, the range of influence, $r_i$, defined as the distance from a given tree to the nearest tree having a larger diameter. Trees compete for light mostly with individuals of their size or greater, and so $r_i$ can be considered as the distance to the nearest significant competitor. The PDF of $r_i$ is predicted to be (see Table 1)

$$p_{r_i}(r_i / h_c) = r_i^{-3} f_{r_i}(r_i / h_c^H) \,, \tag{3}$$

with an exponent value of 3 independent of the specific value of *H*. The cut-off dependence arises because $r_i$ is expected to scale isometrically with $r_{cro}$ due to competition for space, implying $r_i \sim r_{cro} \sim h^H$.

Figure 1 shows a plot that confirms the above prediction and, surprisingly, the power law behaviour holds up to the size of the forest that is much larger than the largest crown size and there is little need for the scaling function to provide the expected cut-off of pure power law behaviour. We will utilize an analysis of this quantity, which exhibits pure power law behaviour valid over a wide range, to deduce whether scaling holds in the BCI forest. Before we do that, let us ask why the exponent value is 3, independent of *H*. Assuming a random distribution of trees within the forest, which ought to be true for trees separated by a large distance, one can prove that the PDF of $r_i$ is a power law with the same decay exponent of 3. It is this harmonious matching of exponent values at short length scales (where the tree shape and the width of the crown matter) and the long length scale behaviour (where one may apply random distribution considerations), which leads to an almost perfect power law relationship extending over a wide range of $r_i$.

We now turn to an application of finite size scaling through a powerful scaling collapse procedure (*29,30*). The conditional PDF of $r_i$, the range of influence, given the stem diameter $r$, is predicted to be,

$$P_{r_i}(r_i|r) = \frac{1}{r_i} F_{r_i}\left(\frac{r_i}{r^{\frac{2H}{1+2H}}}\right) \qquad (4).$$

Eq. (4) represents the probabilistic generalization of the deterministic counterpart $r_i \sim r^{2H/(1+2H)}$. The pre-factor $1/r_i$ ensures that the average of $r_i$ scales as $r^{2H/(1+2H)}$ (more generally, the n-th moments of $r_i$ scale as $(r^n)^{2H/(1+2H)}$). The prediction from Eq. (4) is that a plot of the cumulative PDF (which incorporates the pre-factor on the right hand side) versus $r_i / r^{2H/(1+2H)}$ for various size classes, over which the scaling framework holds, must collapse (*29,30*) on to a single plot. Figure 2 shows the collapse plot of $P_{r_i}(r_i|r)$ for the predicted exponent of $H=1$. The collapse is optimal for trees with diameters in the range of 2.4 - 31.8 cm indicating that this is the correct range in which power law behaviour ought to be observed and measured. The collapse begins to break down for larger diameters due to resource limitations. The finite size scaling collapse is an objective method for estimating exponents when pure power law behaviour is not observed over a significant range of values of the variables being studied leading to a spread of exponent values. Figure 3 shows the results of two other tests of the consistency of the scaling theory and are compared with previous predictions of metabolic ecology (*4, 6, 7, 9, 12, 14, 15, 18, 19, 22, 23*). Figure 1S in Supporting Information further demonstrates the validity of our hypothesis.
The scaling framework presented here as well as exponent relationships and exponent values are realized in the self-similar forest model presented below.

**Analytically solvable forest model**

The scaling results hold exactly as predicted by our approach for a simple self-similar model that satisfies all the hypotheses. In two (three) dimensions, the forest is represented by a rectangle $L \times h_c$ (a volume $L \times L \times h_c$), where $L$ and $h_c$ represent the linear size of the forest and the height of the tallest trees respectively with $L \gg h_c$. At the zero-th step, we start to fill the ecosystem with the highest trees represented by triangles of height $h_c$ and crown extension $h_c$ with $H=1$ (upside down pyramids of height $h_c$ and square base $h_c \times h_c$). The area (volume) occupied by a single large tree is a measure of the metabolic rate, $B = h_c^2/2$ ($B \sim h_c^3$). The number of the tallest trees is $\rho = L/h_c$ ($\rho = (L/h_c)^2$). In the next steps - labelled with index $t>0$ - we introduce trees of a different shape: in two dimensions, the tree now has the shape of a rhombus of height $h(t) = h_c/2^{t-1}$ and transverse extension $h_c/2^t$. As shown in Figure 4, they perfectly fill in the empty spaces between trees in the former levels. (In three dimensions, the shape of the tree is an upside down pyramid with a square base with height $h(t) \sim h_c/2^{t-1}$ and base side $\sim h_c/2^t$. A crown of variable shape is attached to the base so that the space between pyramids of two consecutive levels is completely filled, and each tree occupies the same volume proportional to $h(t)^3$. The three dimensional case is self-similar as well but is harder to visualize).

In two dimensions, the metabolic rate of a tree of level $t$ is $B(t)=(h_c/2^t)^2$ and there are exactly $N(t)=\rho 2^{t-1}$ trees in step $t$ and $N_>(t)=\rho 2^t$ total trees at step $t$. Thus the total number of trees with metabolic rate larger than B is $N_>(B) = \dfrac{L\Theta(h_c^2/2 - B)}{B^{1/2}}$, where $\Theta(x)$ is the step function equal to $1$ when $x > 0$ and $0$ otherwise. The PDF is given by $p_B(B|h_c) \propto -\dfrac{d}{dB}N_>(B) \propto \dfrac{\Theta(h_c^2/2 - B)}{B^{3/2}}$, yielding a probability density function of the form $p_B(B|h_c) = B^{-\varphi} f_B(B/h_c^2)$ with a scaling function $f_B(B/h_c^2) \propto \Theta\left[1 - \dfrac{2B}{h_c^2}\right]$.

The cut-off arises because the largest tree in the model has an area and thus a metabolic rate equal to $h_c^2/2$. Thus the power law $p_B(B) \sim B^{-\varphi}$ holds only when $B < h_c^2/2$ – the range over which scaling is observed grows as $h_c$ increases. The other PDFs can be found in a similar manner. For example, $p_h(h|h_c) \propto \dfrac{\Theta(h_c - h)}{h^2}$ which is consistent with the PDF for B because $B \sim h^2$. Note also that trees of the same size are uniformly distributed, a postulate used to derive the $d_s$ scaling shown in Figure 3b. A similar analysis in the realistic three dimensional case yields $p_B(B|h_c) \propto \frac{1}{B^{5/3}} \Theta\left(1 - \frac{cB}{h_c^3}\right)$, where $c$ is a numerical constant. The exponent value of 5/3 is in accord with the results presented in Table 1. Other scaling laws and PDFs follow in a straightforward manner. Using Hypothesis 2, $B \sim r^2$, the above equation gives for the diameter PDF: $p(r|h_c) \propto 1/r^{7/3} \Theta(1 - cr/h_c^{3/2})$. On introducing more realistic ingredients such as randomness in the plant position and size, one finds that the exponent of the power law is robust and does not change whereas the $\Theta$ function becomes a smooth function, $f_r(r/h_c^{3/2})$, with the characteristics described in hypothesis 4.

**Discussion**

We have demonstrated that scaling (*29,30*) provides a powerful framework for the analysis of forest data even in the absence of power law behaviour over extended scales and yields predictions in accord with data. We have shown that seemingly distinct patterns are all derivable from a single tree shape exponent, *H*, thus predicting links between them. The scaling results hold exactly as predicted by our approach for an exactly solvable self-similar model which satisfies all the hypotheses. For tropical forests, we have found that the maximum value of *H=1*, corresponding to the optimization of tree metabolic rate, provides a good fit to data. Our framework is eminently suited for the study of forests across the globe (when detailed information pertaining to the locations of trees and their diameters, heights, and crown shape become available) to understand the steady-state conditions allowing the maximal use of resources, to elucidate the dependence of the value of the shape exponent on latitude and climate, and to understand the effect of disturbances on forest structure, carbon stock and sink.

**Acknowledgments**:

We are indebted to Jim Brown and his collaborators for the stimulating ideas presented in their papers on the subject. The BCI forest dynamics research project was made possible by National Science Foundation grants to Stephen P. Hubbell, support from the Center for Tropical Forest Science, the Smithsonian Tropical Research Institute, the John D. and Catherine T. MacArthur Foundation, the Mellon Foundation, and the Celera Foundation. A.M. acknowledges the support of Fondazione Cariparo – Padova.

**Footnotes**:

∗ This assumption follows from the observation that the leaves are supplied nutrients at a rate independent of tree height, facilitated by xylem tapering.

**References**

1. M. Kleiber, *The Fire of Life: An Introduction to Animal Energetics*. (Wiley, New York, 1961).

2. K. Schmidt-Nielsen, *Scaling*. (Cambridge Univ. Press, 1984).

3. Pacala, S.W., Canham, C.D., & Silander Jr, J.A., Forest models defined by field measurements: I. The design of a northeastern forest simulator. *Canadian Journal Forest Research* **23**, 1980-1988 (1993).

4. West, G.B., Brown, J.H., & Enquist, B.J., A general model for the origin of allometric scaling laws in biology. *Science* **276**, 122-126 (1997).

5. Banavar, J.R., Maritan, A., & Rinaldo, A., Size and form in efficient transportation networks. *Nature* **399**, 130-132 (1999).

6. Enquist, B.J., Brown, J.H., & West, G.B., Allometric scaling of plant energetics and population density. *Nature* **395**, 163-165 (1998).

7. West, G.B., Brown, J.H., & Enquist, B.J., A general model for the structure and allometry of plant vascular systems. *Nature* **400**, 664-667 (1999).

8. Damuth, J., Scaling of growth: Plants and animals are not so different. *Proceedings of the National Academy of Sciences of the United States of America* **98**, 2113-2114 (2001).

9. Enquist, B.J. & Niklas, K.J., Invariant scaling relations across tree-dominated communities. *Nature* **410**, 655-660 (2001).

10. S. P. Hubbell, *The Unified Neutral Theory of Biodiversity and Biogeography*. (Princeton Univ. Press, 2001).

11. Meinzer, F.C., Goldstein, G., & Andrade, J.L., Regulation of water flux through tropical forest canopy trees: Do universal rules apply? *Tree Physiology* **21**, 19-26 (2001).

12. Niklas, K.J. & Enquist, B.J., Invariant scaling relationships for interspecific plant biomass production rates and body size. *Proceedings of the National Academy of Sciences of the United States of America* **98**, 2922-2927 (2001).

13. Allen, A.P., Brown, J.H., & Gillooly, J.F., Global biodiversity, biochemical kinetics, and the energetic-equivalence rule. *Science* **297**, 1545-1548 (2002).

14. Enquist, B.J., Universal scaling in tree and vascular plant allometry: toward a general quantitative theory linking plant form and function from cells to ecosystems. *Tree Physiology* **22**, 1045-1064 (2002).

15. Enquist, B.J. & Niklas, K.J., Global allocation rules for patterns of biomass partitioning in seed plants. *Science* **295**, 1517-1520 (2002).

16. Coomes, D.A., Duncan, R.P., Allen, R.B., & Truscott, J., Disturbances prevent stem size-density distributions in natural forests from following scaling relationships. *Ecology Letters* **6**, 980-989 (2003).

17. Kohyama, T., Suzuki, E., Partomihardjo, T., Yamada, T., & Kubo, T., Tree species differentiation in growth, recruitment and allometry in relation to maximum height in a Bornean mixed dipterocarp forest. *Journal of Ecology* **91**, 797-806 (2003).

18. Niklas, K.J., Midgley, J.J., & Enquist, B.J., A general model for mass-growth-density relations across tree-dominated communities. *Evolutionary Ecology Research* **5**, 459-468 (2003).

19. Brown, J.H., Gillooly, J.F., Allen, A.P., Savage, V.M., & West, G.B., Toward a metabolic theory of ecology. *Ecology* **85**, 1771-1789 (2004).

20. Muller-Landau, H.C. *et al.*, Testing metabolic ecology theory for allometric scaling of tree size, growth and mortality in tropical forests. *Ecology Letters* **9**, 575-588 (2006).

21. Muller-Landau, H.C. *et al.*, Comparing tropical forest tree size distributions with the predictions of metabolic ecology and equilibrium models. *Ecology Letters* **9**, 589-602 (2006).

22. Enquist, B.J., West, G.B., & Brown, J.H., Extensions and evaluations of a general quantitative theory of forest structure and dynamics. *Proceedings of* *the National Academy of Sciences of the United States of America* **106**, 7046-7051 (2009).

23. West, G.B., Enquist, B.J., & Brown, J.H., A general quantitative theory of forest structure and dynamics. *Proceedings of the National Academy of Sciences of the United States of America* **106**, 7040-7045 (2009).

24. Banavar, J.R., Moses, M.E., Brown, J.H., Damuth, J., Rinaldo, A., Sibly, R.M., & Maritan, A., A general basis for quarter power scaling in biology. (preprint).

25. Hubbell, S.P., Condit, R., & Foster, R.B., Barro Colorado Forest Census Plot Data, Available at http://ctfs.si/edu/datasets/bci, (2005).

26. Kadanoff, L.P., Scaling laws for Ising models near Tc. *Physics* **2**, 263–272 (1966).

27. Widom, B., The critical point and scaling theory. *Physica* **73**, 107-118 (1974).

28. Wilson, K.G., The renormalization group and critical phenomena. *Reviews of Modern Physics* **55**, 583-600 (1983).

29. Fisher, M.E., *Critical Phenomena*. (Academic Press, New York, 1971).

30. Stanley, H.E., Scaling, universality, and renormalization: Three pillars of modern critical phenomena. *Reviews of Modern Physics* **71**, 358-366 (1999).

31. T. Anfodillo, V. Carraro, M. Carrer, C. Fior, S. Rossi, Convergent tapering of xylem conduits in different woody species, *New Phytol.* **169**, 279-290 (2006).

**Figure Captions**

**Figure 1.** Cumulative probability distribution of the range of influence $r_i$ for the BCI dataset (1995) (*25*). $p^>_{r_i}(r_i)$ is the fraction of trees whose minimum distance to a tree of bigger size is greater than $r_i$ (measured in meters). The red solid line is a power law with exponent -2, equivalent to an exponent of -3 for the PDF. Surprisingly, the power law behaviour holds up to the size of the forest that is much larger than the largest crown size and there is little need for the scaling function to provide the expected cut-off of pure power law behaviour. This fact and that this exponent is independent of $H$ has a simple interpretation. Indeed by assuming a random distribution of trees within the forest, which ought to be true for trees separated by a large distance, one can prove that the PDF of $r_i$ is a power law with the same decay exponent of 3. It is this harmonious matching of exponent values at short length scales (where the tree shape and the width of the crown matter) and the long length scale behaviour (where one may apply random distribution considerations), which leads to an almost perfect power law relationship extending over a wide range of $r_i$.

**Figure 2.** **(a)** Scaling collapse plot. The inset shows the probability of having a range of influence $\geq r_i$ for trees with diameter in the interval $(r, r+\delta r)$. We divided the BCI tree-diameter dataset (*25*) in 1cm-size bins and for each bin we calculated the cumulative distribution of distances from the nearest neighbour tree of larger size $P^>_{r_i}(r_i / r)$. These distributions are shown for $r$ ranging between 1.4cm and ~49.4cm in the inset. The main figure depicts a scaling collapse plot of the curves shown in the inset. The scaling framework predicts a collapse plot (see main figure) when the cumulative distributions of distances are plotted against the scaling variable

$r_i/r^{2/3}$ when one is in the scaling regime. The grey curves in the inset do not collapse and the black curves define the range of tree diameters over which scaling is observed (2.4cm to ~31.8cm).

**(b)** Comparison of scaling collapses corresponding to H=1/2 (left) and H=1 (right). These distributions are calculated for sets of trees grouped in 1cm-bins of diameter and are shown for $r$ ranging between 1.4cm and ~49.4cm. H=1 clearly provides a better collapse of the $P_{r_i}^>(r_i/r)$ distributions, allowing one to rule out the exponent value H=1/2. When the scaling function is substantially constant, H=1 leads to $p(r|h_c) \propto 1/r^{7/3}$ whereas $H=1/2$ yields the prediction $p(r|h_c) \propto 1/r^2$.

**Figure 3.** **(a)** The cumulative PDF of tree diameters in the BCI forest (1995) (*25*). $p_r^>(r/r_c)$ is the fraction of trees with diameter greater than or equal to $r$. The black dots correspond to diameters in the interval from ~2.4cm to ~31.8cm (the range over which scaling is expected to hold from the scaling collapse plot in Figure 2). The red solid line indicates our predicted exponent of -4/3 (derived from a power law probability density with exponent -7/3). The blue dashed line depicts the exponent of -1 (corresponding to a probability density with exponent -2) and is shown for comparison.

**(b)** Plot of the average distance from the nearest neighbour individual in the same size class, $d_s$, versus the tree diameter. We divided the BCI tree-diameter dataset (*25*) in 2 cm bin size and for each bin, we calculated the average distance between nearest neighbour trees belonging to the same bin. The red solid line shows the predicted power law behaviour with exponent 7/6, whereas the blue dashed line is a power law with exponent 1. Our prediction follows from the assumption that the trees in a given size class are distributed uniformly across the forest, thus implying that $d_s \sim P_r(r|r_c)^{-1/2} \sim r^{7/6}$.

**Figure 4.** The two dimensional self-similar forest at step $t=3$. The lines denote the boundaries of each tree, and the number inside each tree is the time step of its creation. The space without numbers is filled with higher generation trees. Note that it is not crucial to have complete space filling. For example, removing trees at the 0-th level does not change the results of our scaling analysis.

**Table Caption**

**Table 1.** **Scaling relationships for tropical forests**

Summary of the key predictions of the idealized scaling framework. The second row shows the exponent $\omega$ characterizing the scaling relationships of the form $y \sim x^\omega$, where $x$ is the tree height $h$ and $y = h, r, r_{cro}, \ldots$. The third row presents the $\omega$ values for the idealized case of $H=1$. Our analysis predicts that the PDF of $y$ satisfies the scaling form $p_y(y \mid y_c) = y^{-\alpha} f_y(y/y_c)$ where $y_c \sim h_c^\omega$ and $f_y$ is a suitable scaling function as explained in the Supporting Information. The corresponding value of the exponent $\alpha$ is predicted to be equal to $1 + 2H/\omega$. As an example, the PDF of the distribution of the metabolic rate $B$ is predicted to be

$$p_B(B \mid h_c) = B^{-\frac{1+4H}{1+2H}} f_B(B/h_c^{1+2H}) \mid_{H=1} = B^{-\frac{5}{3}} f_B(B/h_c^3).$$

**Table 1**

|  | $h$ | $r$ | $r_{cro}$ | $r_i$ | $d_s$ | $B$ | $M$ |
|---|---|---|---|---|---|---|---|
| $\omega$ | 1 | $\dfrac{1+2H}{2}$ | $H$ | $H$ | $\dfrac{1+6H}{4}$ | $1+2H$ | $2(1+H)$ |
| $\omega\vert_{H=1}$ | 1 | 3/2 | 1 | 1 | 7/4 | 3 | 4 |
| $\alpha$ | $1+2H$ | $\dfrac{1+6H}{1+2H}$ | 3 | 3 | $\dfrac{1+14H}{1+6H}$ | $\dfrac{1+4H}{1+2H}$ | $\dfrac{1+2H}{1+H}$ |
| $\alpha\vert_{H=1}$ | 3 | 7/3 | 3 | 3 | 15/7 | 5/3 | 3/2 |

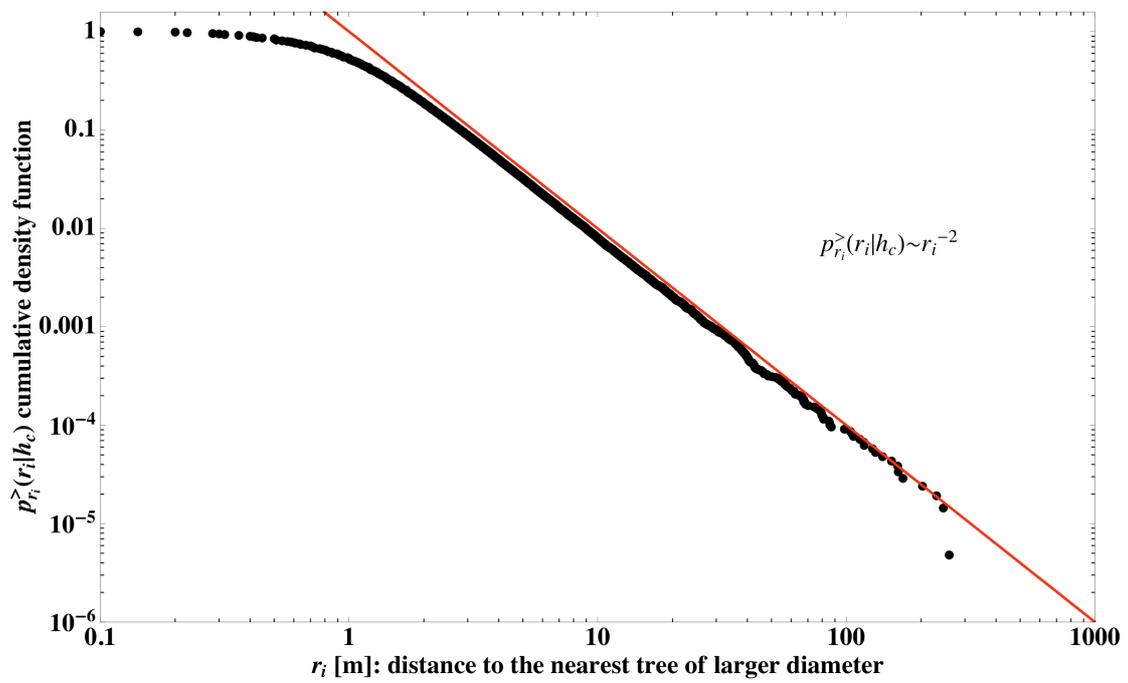

**Figure 1**

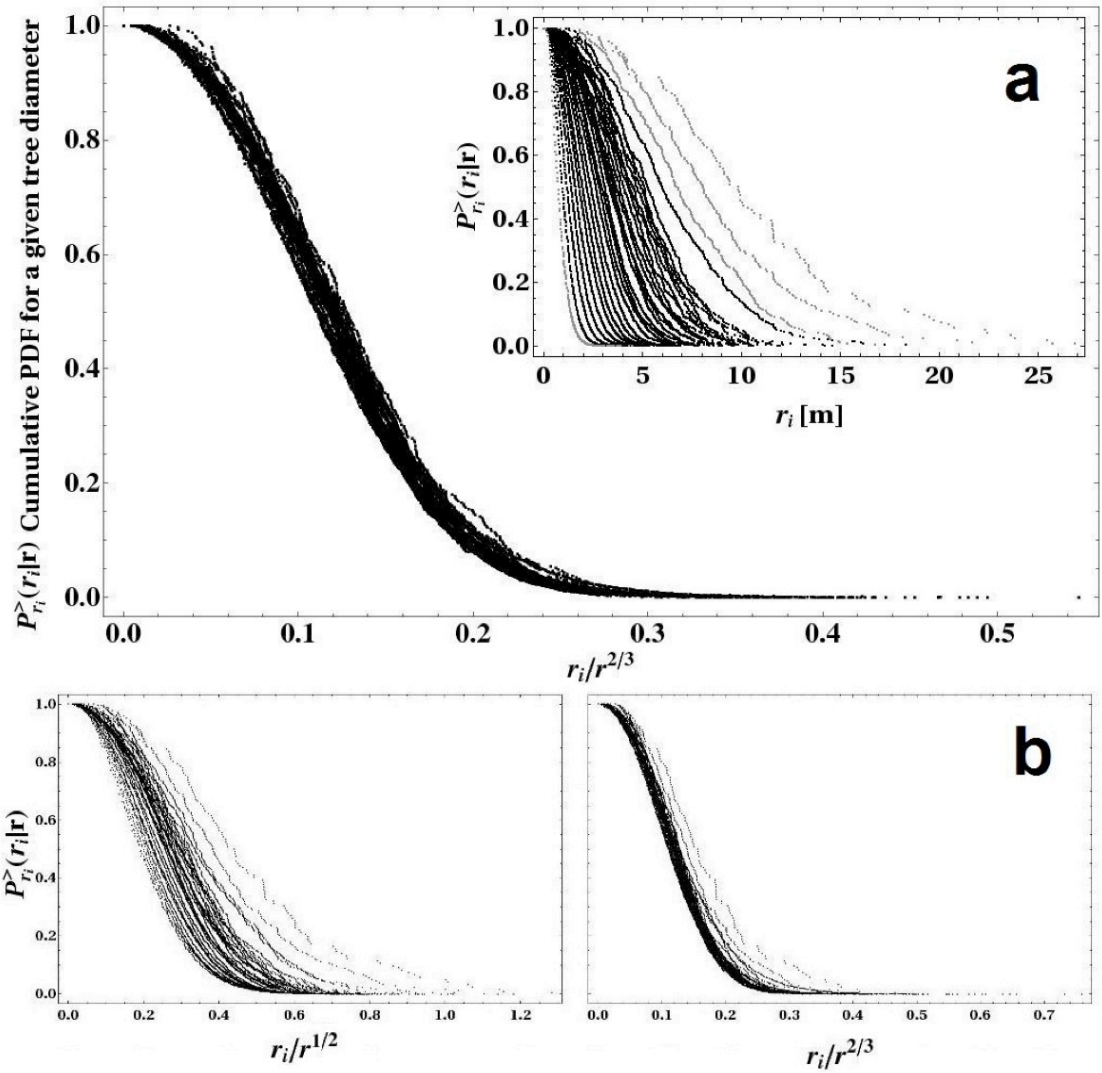

**Figure 2**

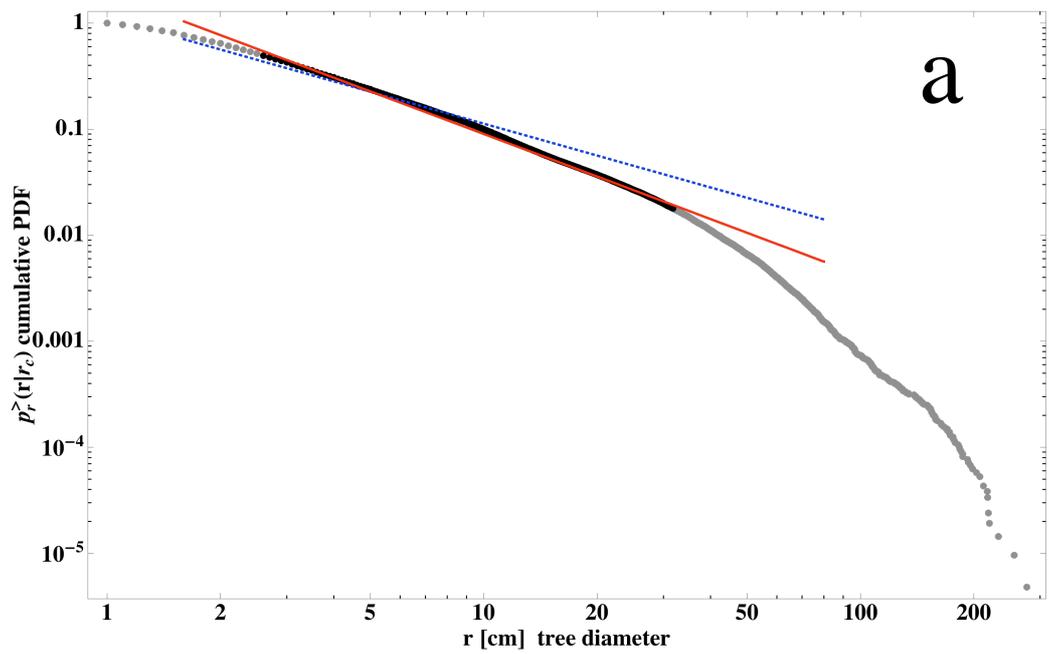
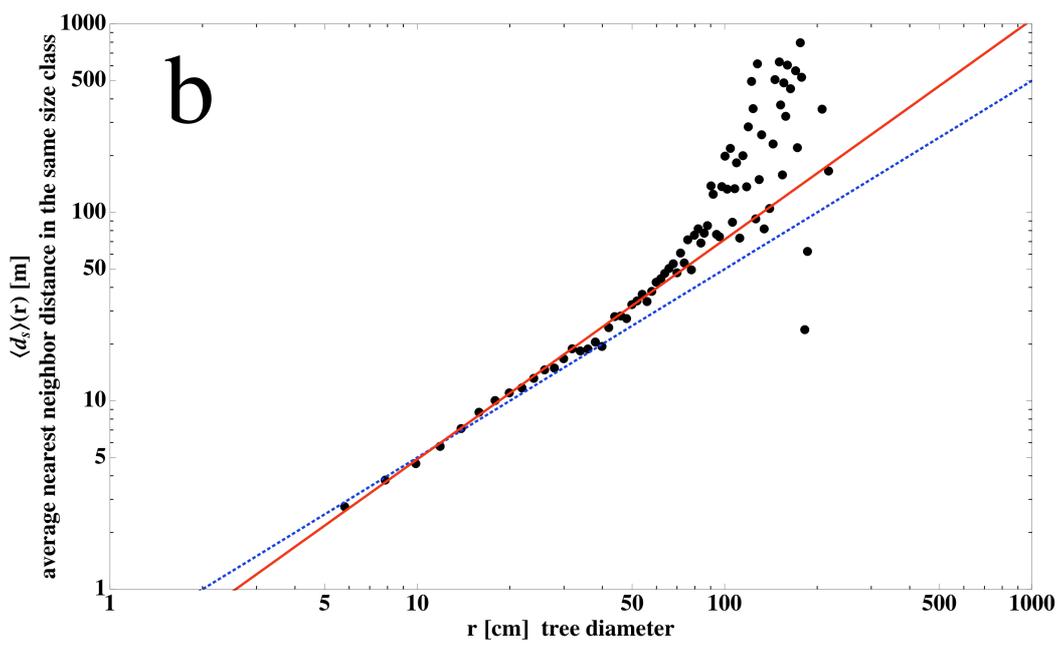

**Figure 3**

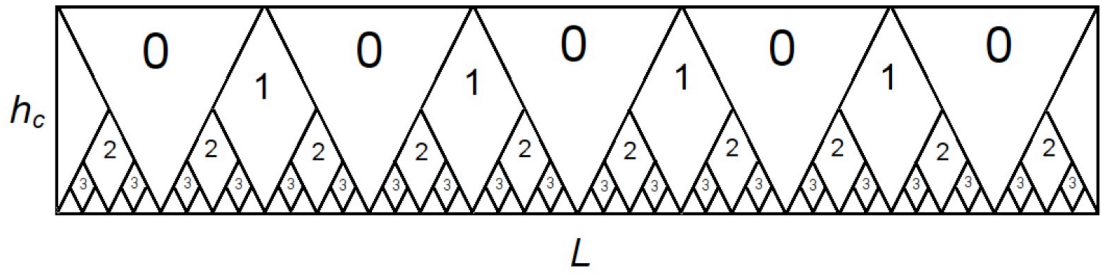

**Figure 4**

**SUPPORTING INFORMATION**

**Self-Similarity and Scaling in Forest Communities**

Filippo Simini[1], Tommaso Anfodillo[2], Marco Carrer[2], Jayanth R. Banavar[3],

& Amos Maritan[1]

[1] *Dipartimento di Fisica ``G. Galilei'', Università di Padova, CNISM and INFN, via Marzolo 8, 35131 Padova, Italy.*

[2] *Dipartimento Territorio e Sistemi Agro-Forestali, Università di Padova, AGRIPOLIS, viale dell'Università 16, 35020 Legnaro (PD), Italy.*

[3] *Department of Physics, The Pennsylvania State University, 104 Davey Laboratory, University Park, Pennsylvania 16802.*

**Contents**

- Scaling and self-similarity.
- Derivation of exponent $\alpha$ for height PDF using finite scaling.
- Energy equivalence.
- Link between metabolic ecology and demographic equilibrium theory.
- Derivation of results presented in Table 1.
- Confirmation of the validity of finite size scaling.

**Scaling and self-similarity**

Scaling and power law relationships are observed when the phenomenon being studied does not exhibit a characteristic length scale (*1-5*). Typically, there are both lower and upper cut-off scales for power law behaviour and if these are well separated (say, by several orders of magnitude), scaling could hold in an intermediate range. In physical systems, one can discern the scaling regime by increasing the upper cut-off scale or the correlation length by adjusting the temperature or the pressure closer to its critical value. No such tuning is possible in an ecological community. The diameter distribution of trees in a forest has a lower cut-off scale set by the size of the plant upon recruitment whereas the upper cut-off is necessarily less than the typical diameter of the largest tree in the forest. Determining the scaling regime and even verifying that a scaling description is valid in an ecological community can be a challenge.

Finite size scaling postulates that the PDF of tree diameters has the form $p_r(r|r_c) = r^{-\alpha} f_r(r/r_c)$ when $r$ is larger than a lower (unspecified) cut-off value. This scaling form is a power law decay $r^{-\alpha}$ characterized by an exponent $\alpha$, but modified by a scaling function $f_r(r/r_c)$, where $r_c$ represents the upper cut-off. Over a range of $r$ values, for which the scaling function is approximately constant, one obtains pure power law behaviour. The scaling function $f_r(r/r_c)$ has the property that it decays to zero rapidly when its argument $(r/r_c)$ becomes larger than 1 or when the tree diameter becomes larger than the cut-off value. In this regime, the PDF is dominated by a characteristic length, the cut-off scale, and pure power law behaviour is lost. This ensures that the PDF appropriately vanishes when the tree diameter becomes larger than its cut-off value. Indeed, power law scaling is expected to hold only when the diameter is much smaller than its cut-off value. The exponent $\alpha$ is expected to be universal and depends only on certain essential attributes, whereas the scaling function $f_r$ can depend on details such as the climate and the resource availability in a given forest. For another variable, such as the height $h$ or the crown radius $r_{cro}$, the exponent $\alpha$ of the PDF has to be determined using the transformation rules described below. Such a scaling form can be used to describe the PDF of variables beyond the range over which they exhibit pure power law behaviour.

When a scaling relation, $y \sim x^{\omega}$, exists between two random variables $x$ and y (for example $x$ could be the height, $h$, and $y$ could be the tree diameter, $r$), it is meant that the conditional probability distribution of $y$ given $x$, $P_y(y|x)$, satisfies the relationship (we use $P$ for the conditional probability of two random variables whereas we use $p$ for PDF of a single random variable):

$$P_y(y|x) = 1/y \; F_y(y/x^{\omega}) \;. \tag{1S}$$

This is the correct generalization of the deterministic relation $y = x^{\omega}$ to a more general case in which the n-th moment scales as $\langle y^n \rangle = c_n x^{\omega n}$, where the $c_n$s are constants. The deterministic case is obtained when $c_n = 1$. In the case finite size scaling holds for the PDF of random variable $x$, $p_x(x|x_c) = x^{-\alpha} f_x(x/x_c)$, the corresponding PDF for the random variable $y$, obeying eq. (1S), is $p_y(y|y_c) = y^{-(\alpha+\omega-1)/\omega} f_y(y/y_c)$ with the cut-offs transforming in the natural manner, i.e. $y_c = x_c^{\omega}$, and two scaling functions $f_x$ and $f_y$ related through an integral equation involving the $F_y$ which appears in eq. (1S). The power law exponent is the one it would expected by the standard change of variable rule for PDF, i.e. $p_y(y) = p_x(x)|dx/dy|$ with $|dx/dy| = y^{(1-\omega)/\omega}/\omega$. As $x$ varies, in principle, one would obtain independent curves $P_y(y|x)$ versus $y$. However, if Eq. (1S) holds, all these curves can be collapsed on to a single curve if one plots $yP_y(y|x)$ (or equivalently the cumulative $P_y^>(y|x) = \int_y^{\infty} dy' P(y'|x)$ ) versus $y/x^{\omega}$. In other words, for a given $x$, the characteristic scale of $y$ is $x^{\omega}$. Viewed in this manner, all curves appear the same. An example where Eq. (1S) holds is shown in Fig. 2.

**Derivation of exponent $\alpha$ for height PDF using finite scaling**

This is a derivation of the scaling exponent $\alpha$ for the PDF of the tree height, which is postulated to have the scaling form $p_h(h|h_c) = h^{-\alpha} f_h(h/h_c)$. We determine the exponent $\alpha$ by obtaining two different measures of the total volume of the forest and equating them. The first, $Ah_c$, is simply the product of the area of the forest and the characteristic height. The second measure is obtained by noting that the total number of trees with height in the range $(h, h+dh)$ is, according to hypotheses 3 and 4, given by $Ap_h(h|h_c)dh$, and the volume of a tree of height $h$ is given by $h^{1+2H}$. Thus, the total volume occupied by the forest in a plot of area $A$ is given by:

$$Ah_c \propto A \int h^{1+2H} p_h(h/h_c) dh \propto h_c^{2+2H-\alpha} A. \tag{2S}$$

The last term is obtained using the assumed scaling form of $p_h(h|h_c)$, making the change of variable $x=h/h_c$, and including into the proportionality constant the integral $\int dx\, x^{1+2H} f_h(x)$. Equating the powers of $h_c$ in the first and third terms of the above equation, one finds that $\alpha = 1+2H$, or equivalently $p_h(h|h_c) \sim 1/h^{1+2H}$ for tree heights $h << h_c$.

**Energy equivalence**

The total energy resources used in the forest is proportional to the total volume of the forest, $Ah_c$. This general result is independent of any scaling assumptions or the value of the $H$ exponent. Thus the characteristic height of trees, $h_c$, is a measure of the average energy use per unit area or equivalently per tree. Therefore if we express the energy equivalence wrt a generic size variable $x \sim h^\gamma$, i.e. $p_x(x) \sim x^{-\frac{1+2H}{\gamma}}$ for $x < x_c \sim h_c^\gamma$, we obtain $E_{tot} \propto A \int x^{\frac{1+2H}{\gamma}} p_x(x|h_c^\gamma) dx \propto A\, h_c^\gamma$ and this is in contrast with the scaling of energy per tree derived above, unless $\gamma=1$.

**Link between metabolic ecology and demographic equilibrium theory**

Our approach allows one to bridge metabolic ecology (*6-18*) with demographic equilibrium theory (*19-21*). On using the ontogenetic growth equation (*22-24*) with the generic finite size scaling assumption (*2, 5*), one gets the growth rate, $g(r) \sim r^c G(r/r_c)$ with $c=(2H-1)/(2H+1)$. The mortality rate, $m(r)$, can be obtained, following demographic equilibrium theory, and requiring that

$$\frac{\partial}{\partial r}\left[g(r) p_r(r|r_c)\right] + m(r) p_r(r|r_c) = 0 \qquad (3S)$$

Inserting the finite size scaling equation for $p_r(r|r_c)$ in the previous equation leads to $m(r) \sim r^b M(r/r_c)$, with $b=c-1= -2/(1+2H)$ and $M(x)=2G(x) - xdG(x)/dx - G(x)(x/f_r(x))(df_r(x)/dx)$. The choice $G, M = constant$ corresponds to the Muller-Landau et al. (*25*) postulate of pure power law behaviour, which is what happens here in the regime $r<r_c$, yielding $M/G=2$, independent of $H$. In ref.(*25*), this ratio was predicted to be 5/3 based on the previous incorrect results of metabolic theory of ref. (*8, 15*). When $H=1$, $c=1/3$ and $b=-2/3$ and are consistent with empirical data (*16*) and agree with the predictions of ref. (*8, 15*).

**Derivation of results presented in Table 1**

Here we present, as an example, the derivation of the probability distribution of metabolic rate *B* given the PDF of tree heights, as in Hypothesis 4, $p_h(h|h_c) = h^{-(1+2H)} f_h(h/h_c)$. As explained earlier, the scaling relation between *B* and $h$, $B \sim h^{1+2H}$, means that the conditional probability $P_B(B|h) = 1/B \, F_B(B/h^{1+2H})$. According to standard rules of combining probabilities one has

$$p_B(B \mid h_c) = \int dh P_B(B \mid h) p_h(h \mid h_c) = \int dh \frac{1}{B} F_B(B/h^{1+2H}) h^{-(1+2H)} f_h(h/h_c) =$$

$$= B^{-\frac{1+4H}{1+2H}} f_B(B/h_c^{1+2H}).$$

After performing the change of variable $z = B/h^{1+2H}$ in the last integral above, we find

$$f_B(x) = \frac{1}{1+2H} \int dz \, z^{-\frac{1+4H}{1+2H}} f_h[(xz)^{1/(1+2H)}] \, F_B(z) \quad .$$

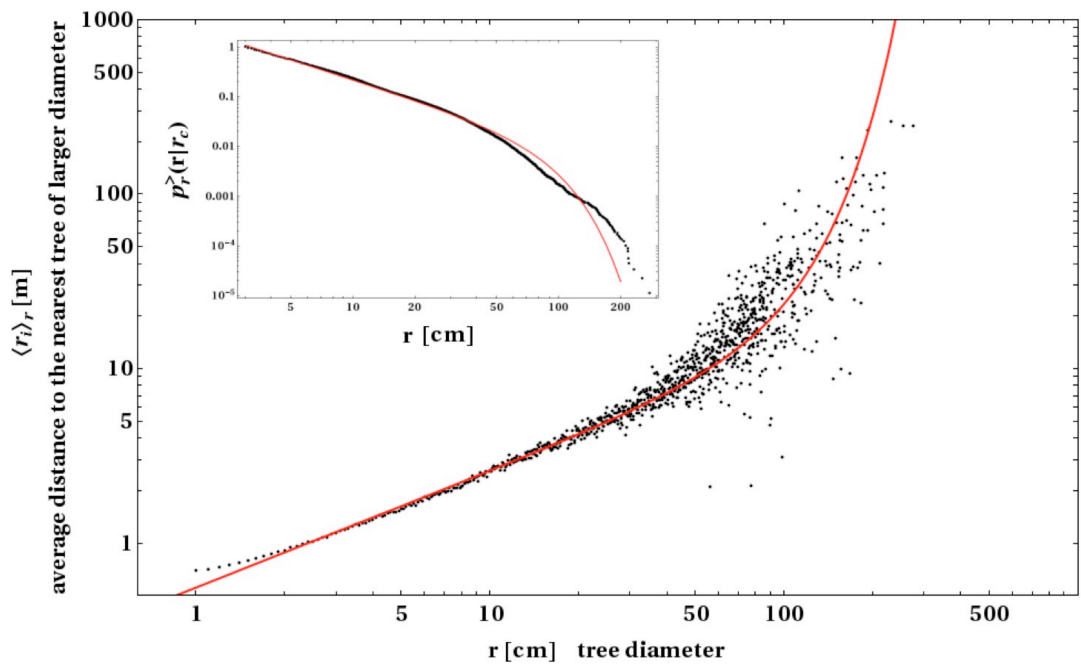

**<u>Figure 1S</u>**

**Confirmation of the validity of finite-size scaling (Figure 1S)**

The main panel of Figure 1S shows a plot of the mean range of influence versus the stem diameter. The BCI diameter dataset were divided in 1 mm-bin classes and for each class we measured the average distance to the nearest larger tree. The inset depicts the cumulative PDF of tree diameters for the BCI dataset (1995)(*26*). $p_r^>(r|r_c)$ is the fraction of trees with diameter greater than or equal to *r*. The solid line shows a least squares fit with the function $p_r^>(r|r_c) = r^{-4/3} g(r/r_c)$, with $g(x) = \exp(-x^2/2)$. This functional form is consistent with our theory with a simple choice of a scaling function with $r_c$, the fitting parameter, equal to 86.6 cm. The precise fitting function *g* is somewhat arbitrary and we have chosen it to be a Gaussian. This choice is based on simplicity and merely serves to demonstrate how scaling can be used to interrelate two distinct patterns. $r_c$ is the cut-off diameter and scaling is expected to hold for length scales much less than this cut-off value. This result is in accord with the direct determination of the scaling range (~2.4cm to ~31.8cm) obtained with the scaling collapse of $P_{r_i}^>(r_i|r)$ (see Figure 2). The solid line in the main figure shows the average range of influence of trees of a given stem diameter, $\langle r_i \rangle_r$, using the estimate $\langle r_i \rangle_r \propto 1/\sqrt{p_r^>(r|r_c)} = 1.2 \ r^{2/3} g(r/r_c)^{-1/2}$ with exactly the same *g(x)* used in the inset and the value of $r_c$ determined therein. The quality of the fit again demonstrates the validity of the scaling framework.

**References**

1. Kadanoff L. P., Scaling laws for Ising models near Tc. *Physics* **2**, 263–272 (1966).

2. Fisher M. E., *Critical Phenomena*. ed. Green M.S. (Acad. Press, New York, 1971).

3. Widom B., The critical point and scaling theory. *Physica* **73**, 107-118 (1974).

4. Wilson K. G., The renormalization group and critical phenomena. *Rev. Mod. Phys.* **55**, 583-600 (1983).

5. Stanley H. E., Scaling, universality, and renormalization: Three pillars of modern critical phenomena. *Rev. Mod. Phys.* **71**, 358-366 (1999).

6. West G. B., Brown J. H., Enquist B. J., A General Model for the Origin of Allometric Scaling Laws in Biology. *Science* **276**, 122-126 (1997).

7. Enquist B. J., Brown J. H., West G. B., Allometric scaling of plant energetics and population density. *Nature* **395**, 163-165 (1998).

8. Enquist B. J., West G. B., Charnov E. L., Brown J. H., Allometric scaling of production and life-history variation in vascular plants. *Nature* **401**, 907-911 (1999).

9. West G. B., Brown J. H., Enquist B. J., A general model for the structure and allometry of plant vascular systems. *Nature* **400**, 664-667 (1999).

10. Enquist B. J., Niklas K. J., Invariant scaling relations across tree-dominated communities. *Nature* **410**, 655-660 (2001).

11. Niklas K. J., Enquist B. J., Invariant scaling relationships for interspecific plant biomass production rates and body size. *Proc. Natl. Acad. Sci. U.S.A.* **98**, 2922-2927 (2001).

12. Enquist B. J., Universal scaling in tree and vascular plant allometry: toward a general quantitative theory linking plant form and function from cells to ecosystems. *Tree Physiol.* **22**, 1045–1064 (2002).

13. Enquist B. J., Niklas K. J., Global allocation rules for patterns of biomass partitioning in seed plants. *Science* **295**, 1517-1520 (2002).

14. Niklas K. J., Midgley J. J., Enquist B. J., A general model for mass-growth-density relations across tree-dominated communities. *Evol. Ecol. Res.* **5**, 459-468 (2003).

15. Brown J. H., Gillooly J. F., Allen A. P., Savage V. M., West G. B., Toward a metabolic theory of ecology. *Ecology* **85**, 1771-1789 (2004).

16. Enquist B. J., West G. B., Brown J. H., Extensions and evaluations of a general quantitative theory of forest structure and dynamics. *Proc. Natl. Acad. Sci. U.S.A.* **106**, 7046-7051 (2009).

17. West G. B., Enquist B. J., Brown J. H., A general quantitative theory of forest structure and dynamics. *Proc. Natl. Acad. Sci. U.S.A.* **106**, 7040-7045 (2009).

18. West G. B., Brown J. H., Life's universal scaling laws. *Phys. Today* **57**, 36-42 (2004).

19. Coomes D. A., Duncan R. P., Allen R. B., Truscott J., Disturbances prevent stem size-density distributions in natural forests from following scaling relationships. *Ecol. Lett.* **6**, 980-989 (2003).

20. Kohyama T., Suzuki E., Partomihardjo T., Yamada T., Kubo T., Tree species differentiation in growth, recruitment and allometry in relation to maximum height in a Bornean mixed dipterocarp forest. *J. Ecol.* **91**, 797-806 (2003).

21. Muller-Landau H. C. *et al.*, Comparing tropical forest tree size distributions with the predictions of metabolic ecology and equilibrium models. *Ecol. Lett.* **9**, 589-602 (2006).

22. West G. B., Brown J. H., Enquist B. J., A general model for ontogenetic growth. *Nature* **413**, 628-631 (2001).

23. Banavar J. R., Damuth J., Maritan A., Rinaldo A., Ontogenetic growth (Communication arising) Modelling universality and scaling. *Nature* **420**, 626-626 (2002).

24. Stegen J. C., White E. P., On the relationship between mass and diameter distributions in tree communities. *Ecol. Lett.* **11**, 1287-1293 (2008).

25. Muller-Landau H. C. *et al.*, Testing metabolic ecology theory for allometric scaling of tree size, growth and mortality in tropical forests. *Ecol. Lett.* **9**, 575-588 (2006).

26. Hubbell S. P., Condit R., Foster R. B., *Barro Colorado Forest Census Plot Data*. Url: http://ctfs.si/edu/datasets/bci (2005).